\documentclass[aps,prl,twocolumn,floatfix]{revtex4-1}
\usepackage{graphicx}
\usepackage{dcolumn}
\usepackage{bm}
\usepackage{ amssymb }

     \usepackage{xcolor} \usepackage{ulem} \usepackage{changebar}

\definecolor{ins}{rgb}{0, 0.5, 0}

\usepackage{array}
\usepackage{subfigure}
\usepackage{epstopdf}
\graphicspath{{Figures/}}

\def\gtwid{\mathrel{\raise.3ex\hbox{$>$\kern-.75em\lower1ex\hbox{$\sim$}}}}
\def\alt{\mathrel{\raise.3ex\hbox{$<$\kern-.75em\lower1ex\hbox{$\sim$}}}}

\def\agt{\mathrel{\raise.3ex\hbox{$>$\kern-.75em\lower1ex\hbox{$\sim$}}}}
\def\alt{\mathrel{\raise.3ex\hbox{$<$\kern-.75em\lower1ex\hbox{$\sim$}}}}

\newcommand{\be}{\begin{equation}}
\newcommand{\ee}{\end{equation}}

\newcommand\Pra{\mbox{\textrm{Pr}}} 
\newcommand\Ra{\mbox{\textrm{Ra}}} 

\begin{document}

\title{Logarithmic Spatial Variations and Universal $f^{-1}$ Power Spectra of Temperature Fluctuations  in Turbulent Rayleigh-B\'enard Convection}

\author{Xiaozhou He,$^{1}$ Dennis P. M. van Gils,$^{1}$ Eberhard Bodenschatz,$^{1,2,3}$ and Guenter Ahlers,$^{1,4}$\\
{\textrm{(International Collaboration for Turbulence Research)}}\\
}

\affiliation{\vskip 12pt
$^1$Max Planck Institute for Dynamics and Self Organization, D-37073 G\"ottingen, Germany%
\\
$^{2}$Institute for Nonlinear Dynamics, University of G\"ottingen, D-37073 G\"ottingen, Germany
\\
$^{3}$Laboratory of Atomic and Solid-State Physics and Sibley School of Mechanical and Aerospace Engineering, Cornell University, Ithaca, New York 14853
\\
$^4$Department of Physics, University of California, Santa Barbara, California 93106, USA%
\\
}

\date{\today}

\begin{abstract}
We report measurements of the temperature variance $\sigma^2(z,r)$ and frequency power spectrum $P(f,z,r)$ ($z$ is the distance from the sample bottom and $r$ the radial coordinate) in turbulent Rayleigh-B\'enard convection (RBC) for Rayleigh numbers $\Ra = 1.6\times10^{13}$ and $1.1\times10^{15}$ and for a Prandtl number $\Pra \simeq 0.8$ for a sample with a height $L = 224$ cm and aspect ratio $D/L = 0.50$ ($D$ is the diameter). For $z/L \alt 0.1$~ $\sigma^2(z,r)$ was consistent with a logarithmic dependence on $z$, and there was a universal (independent of \Ra, $r$, and $z$) normalized spectrum which, for $0.02 \alt f\tau_0 \alt 0.2$, had the form $P(f\tau_0) = P_0 (f\tau_0)^{-1}$ with $P_0 =0.208 \pm 0.008$ a universal constant.  
Here $\tau_0 = \sqrt{2R}$ where $R$ is the radius of curvature of the temperature autocorrelation function $C(\tau)$ at $\tau = 0$. For $z/L \simeq 0.5$ the measurements yielded $P(f\tau_0) \sim (f\tau_0)^{-\alpha}$ with $\alpha$ in the range from 3/2 to 5/3. All the results are similar to those for velocity fluctuations in shear flows at sufficiently large Reynolds numbers, suggesting the possibility of an analogy between the flows that is yet to be determined in detail.
\end{abstract}


\maketitle

Turbulent thermal convection is an important phenomenon in many natural processes, for instance in climatology \cite{DDSC00}, oceanography \cite{MS99}, geophysics \cite{CO94}, astrophysics \cite{Bu94}, and industry. In experiments, it can be generated for instance in a confined system between two horizontal plates separated by a distance $L$ and heated from below in the presence of gravity. This system is known as Rayleigh-B\'enard convection (RBC) \cite{Ah09, AGL09, LX10, CS12}. 

RBC is frequently studied in a cylindrical sample of height $L$ and diameter $D$. Its properties are determined by the Rayleigh number $\Ra \equiv \alpha g \Delta T L^3 / (\nu \kappa)$, the Prandtl number $\Pra \equiv \nu/\kappa$, and the aspect ratio $\Gamma \equiv D/L$. Here $g$ is the gravitational acceleration, $\Delta T = T_b - T_t$ is the temperature difference between the bottom ($T_b$) and the top ($T_t$) plate, and $\alpha$, $\nu$, and $\kappa$ are, respectively, the thermal expansion coefficient, the kinematic viscosity, and the thermal diffusivity of the fluid. 

When $\Ra$ is not too large (say $\Ra \alt 10^{14}$), there are thin laminar boundary layers (BLs) adjacent to the top and bottom plates with most of the temperature difference sustained by them, while the ``bulk" of the fluid between these BLs is turbulent. The conventional view was that the bulk is nearly isothermal. This state is known as ``classical" RBC. As $\Ra$ increases and exceeds a critical value $\Ra^*$ which for $\Pra \simeq 1$ is  ${\cal O}(10^{14})$, the shear stress from the turbulent bulk  will become sufficiently large to force the BLs into a turbulent state as well and the system enters the ``ultimate" state which is expected to be asymptotic as $\Ra$ tends toward infinity \cite{Kr62, Sp71, GL11, HFNBA12}. 

Recently, it was found that the time-averaged temperature $\langle T(t,z,r)\rangle_t$ ($z$ is the vertical and $r$ the radial coordinate), both in the classical and the ultimate state but outside the BLs, varies logarithmically with the distance $z/L$ from the plates when this distance is not too large (say $z/L \alt 0.1$ or so) \cite{ABFGHLSV12,ABH14}. Similar logarithmic behavior is well known from mean velocity profiles of near-wall turbulence in shear flows, such as pipe, channel, and Taylor-Couette flows \cite{WW89, HVBS12, HSCKLS13}; there it is known as the ``Law of the Wall" \cite{Pr25, Ka30, Pr32} (for recent reviews, see \cite{MMMNSS10, SMM11}). For the ultimate state of RBC a logarithmic dependence had been predicted for $T(z)$ by Grossmann and Lohse \cite{GL11,GL12}. For classical RBC its discovery came as a surprise, but one  theoretical explanation was offered very recently \cite{SCCZBH14}.  

In this Letter we report measurements of the temperature variance $\sigma^2(z,r)$ and of temperature temporal frequency spectra $P(f,z,r)$ in the bulk of RBC for $\Ra \simeq10^{13}$ and $10^{15}$ which are representative of the phenomena observed in both the classical and the ultimate state. Outside the BLs but in the near-wall range $z/L \alt 0.1$, we found that $\sigma^2$ also varied logarithmically with $z/L$. The normalized spectra, when scaled by $\tau_0 = \sqrt{2R}$ where $R$ is the radius of curvature of the time autocorrelation function $C(\tau)$ at $\tau=0$, had a universal form (i.e., were independent of \Ra, $r$, and $z$). For $0.02 \alt f \tau_0 \alt 0.2$ they were well described by  $P(f\tau_0) =P_0 \times (f\tau_0)^{-1}$ with $P_0$ a universal constant equal to  $0.208 \pm 0.008$.  Quite remarkably, within our resolution the same universal constant was found in the classical and the ultimate state. To our knowledge, these findings are not explained by present theories for RBC. Away from the near-wall region, near the horizontal midplane of the sample ($z/L \simeq 0.5$), we found a universal spectrum with a larger negative  exponent close to the Obukhov-Corrsin scaling $P(f\tau_0) \sim (f\tau_0)^{-5/3}$ for a passive scalar in turbulent flows \cite{Ob49, Co51}. As we show below by using the elliptic approximation (EA) of space-time correlation functions \cite{HZ06}, we find that there is a one-to-one correspondence between the frequency and wave-number domains, thus making it plausible that our findings for the  temporal spectra $P(f)$ are closely related to the  spatial spectra $E(k)$.

The apparatus and the procedures were discussed before \cite{AFB09, AHFB12}. The sample was contained in the High Pressure Convection Facility II, a cell of height $L = 2.2$ and diameter $D = 1.1$ m which in turn was located in a very large pressure vessel known as the ``Uboot of G\"ottingen." The 25 m$^3$ volume of the  Uboot and sample were filled with  up to 2000 kg of compressed sulfur hexafluoride (SF$_6$) at pressures up to 19 bars as the fluid. By maintaining the mean temperature $T_m = (T_b + T_t)/2$ close to $21.5^{\circ}$C, $\Pra$ was kept at 0.79 (0.86) near $\Ra = 10^{13}$ ($10^{15}$). All measurements had been done under conditions well approximated by the Oberbeck-Boussinesq equations \cite{HFNBA13}. The sample was leveled relative to gravity to within $10^{-4}$ rad. In addition to two sets of thermistors that were reported already in Ref. \cite{ABFGHLSV12}, we installed $58$ new thermistors at various vertical and radial locations to measure the interior temperatures (see the Supplementary Material \cite{SD}).

\begin{figure}
  \centerline{\includegraphics[width=0.95 \columnwidth]{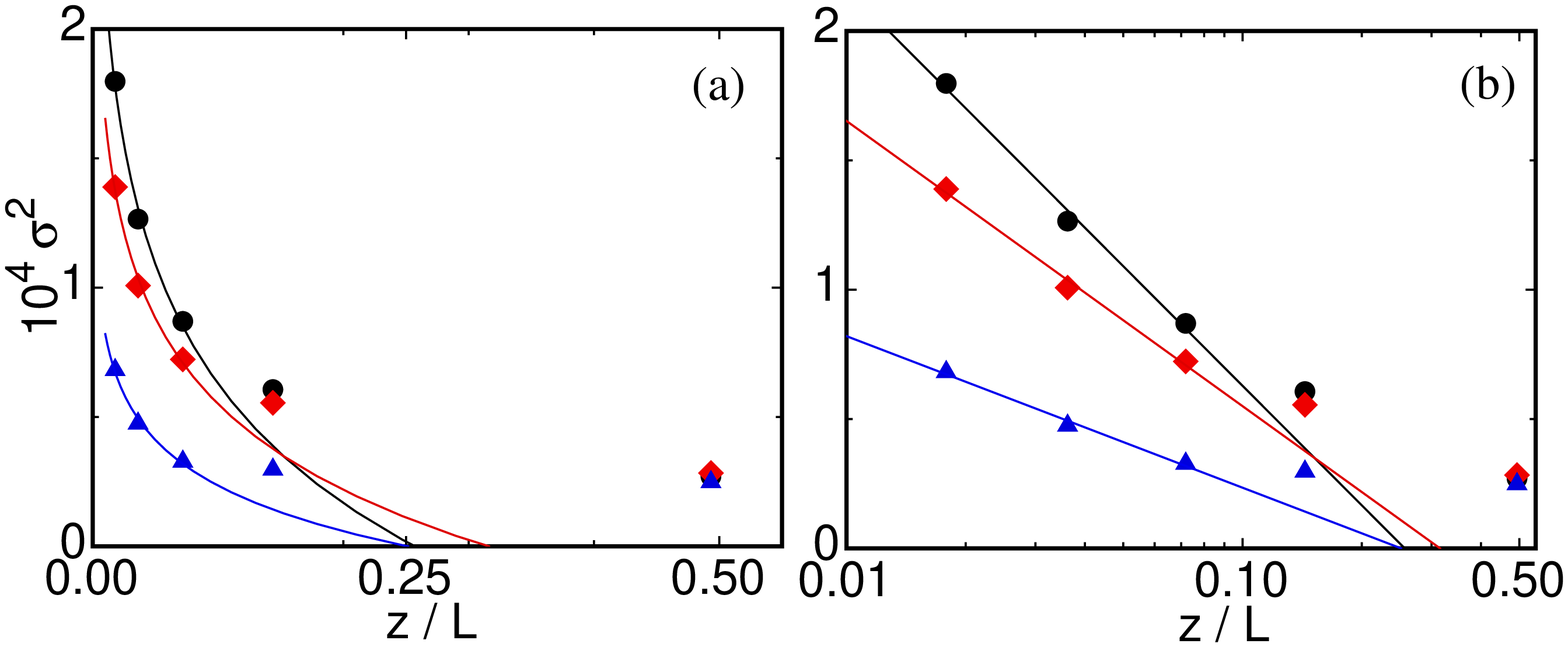}}
\caption{(color online). Experimental results for the variance $\sigma^2(z,r) \equiv \langle [ T(t,z,r) - \langle T(t,z,r) \rangle_t ] ^2 \rangle_t / (T_b - T_t)^2$ as a function of the vertical position $z/L$ for the three radial locations $(R-r)/D = 0.0178$ (black circles), $0.0357$ (red diamonds), and $0.134$ (blue triangles) (a): on a linear  and (b): on a logarithmic horizontal scale. All measurements were for $\Ra=1.08\times10^{15}$ and $\Pra = 0.86$. Equation (\ref{eq:logfitvar}) was fit to the three points with $z/L \leq 0.1$. The solid lines are those fits.
}
\label{fig:log}
\end{figure}

Figure \ref{fig:log} shows vertical profiles of the temperature variance $\sigma^2(z,r) \equiv \langle [ T(t,z,r) - \langle T(t,z,r) \rangle_t ] ^2 \rangle_t / (T_b - T_t)^2$ measured at $\Ra = 1.08 \times 10^{15}$ at three different radial positions $r$. For $z$ well beyond the thermal BLs (which exist for $z/L \ll 10^{-3}$) but $z/L \alt 0.1$, all the $\sigma^2(z,r)$ profiles follow closely a logarithmic dependence on $z/L$, and can be represented by 
\be
\sigma^2(z,r) = M(r)\ln(z/L) + N(r)\ .
\label{eq:logfitvar}
\ee
As the radial location moves closer to the vertical centerline of the sample, $\sigma^2(z,r)$ at small $z/L$ decreases while at midheight ($z/L = 0.5$) it remains the same. This leads to a decease of the amplitude $M(r)$.

\begin{figure}
  \centerline{\includegraphics[width=1\columnwidth]{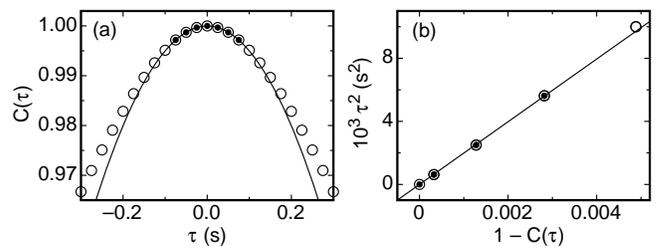}}
  \caption{(a): Temperature autocorrelation function $C(\tau)$ as a function of the time delay $\tau$ measured at $z/L=0.493$ and $(R-r)/D = 0.0178$ for $\Ra=1.08\times10^{15}$. The curve represents the fit of Eq.(\ref{eq:polyfit}) to the solid dots. (b):  $\tau^2$ as a function of the measured values of $1-C(\tau)$ using the data in (a). The line represents the fit of $\tau^2= \tau_0^2 [1-C(\tau)]$ to the solid dots.}
\label{fig:tau0}
\end{figure}

\begin{table}
\caption{Parameters for the temperature spectra shown in Figs.~\ref{fig:P_Ra} and \ref{fig:P_z}, and values of $\tau_0$ used to scale these spectra. The $\tau_0$ values were derived from the temperature autocorrelation functions using Eq.~(\ref{eq:polyfit}).}
\bigskip
\begin{ruledtabular}
\begin{tabular}{ccccc}
\multicolumn{1}{c}{$\Ra$}&
\multicolumn{1}{c}{$z/L$} &
\multicolumn{1}{c}{$(R-r)/D$}&
\multicolumn{1}{c}{$\tau_0$ ( s )}&
\multicolumn{1}{c}{$P_0$}\\
\hline
1.080$\times 10^{15}$&	0.0179&	0.0178&	0.709& 0.202 $\pm$ 0.014\\
1.080$\times 10^{15}$&	0.0362&	0.0178&	0.732& 0.216 $\pm$ 0.014\\
1.080$\times 10^{15}$&	0.0719&	0.0178&	0.801& 0.216 $\pm$ 0.012\\
1.080$\times 10^{15}$&	0.1437&	0.0178&	0.854& \\
1.080$\times 10^{15}$&	0.4933&	0.0178&	1.417& \\
1.080$\times 10^{15}$&	0.0719&	0.0357&	0.829& 0.210 $\pm$ 0.016\\
1.080$\times 10^{15}$&	0.0719&	0.1337&	0.915& 0.196 $\pm$ 0.015\\
1.626$\times 10^{13}$&	0.0179&	0.0178&	1.110& 0.216 $\pm$ 0.012\\
1.626$\times 10^{13}$&	0.4933&	0.0178&	2.051& \\
1.626$\times 10^{13}$&	0.0179&	0.1337&	1.285& 0.202 $\pm$ 0.018\\

\end{tabular}
\end{ruledtabular}
\label{tab:parameters}
\end{table}

As most (but not all \cite{LX10}) previous measurements for RBC, we report the temporal behaviour of thermal fluctuations. Based on the elliptic approximation (EA) of space-time cross-correlation functions \cite{HZ06}, there exists an equivalence between spatial and temporal spectra which is exact to second order in the sense that it is based on a systematic second-order expansion of the correlation functions (see the Supplementary Material \cite{SD}). 

Figure ~\ref{fig:tau0} shows a typical temperature autocorrelation function $C(\tau)$. For $\tau \rightarrow 0$, $C(\tau)$ satisfies
\be
C(\tau) = 1- (\tau / \tau_0)^2\ .
\label{eq:polyfit}
\ee
Fitting Eq. (\ref{eq:polyfit}) to the data with the solid dots ($\tau < 0.1$ s), we get the characteristic time scales $\tau_0$ given in Table~\ref{tab:parameters}. Here the determination of $\tau_0$ from $C(\tau)$ is similar to the determination of the Taylor microscale from velocity space autocorrelation functions in turbulent flows \cite{Po00}.

Elsewhere \cite{HGBA13b} we show that our experimental results for $C(\tau)$ agree well with the EA. An important implication \cite{SD} is that the temperature space correlation function $C(l)$ is equal to $C(\tau)$ provided one chooses $l=V_{eff}\tau$. Here $V_{eff} = \sqrt{U^2+V^2}$ with $U$ and $V$ the mean speed and the root-mean-square velocity, respectively. Similar to $\tau_0$, a typical length scale $\lambda_0$ can be derived from the curvature of $C(l)$, and the EA requires that $\lambda_0 = V_{eff}\tau_0$. Therefore the two scaled correlation functions $C(\tau)$ vs $\tau/\tau_0$ and $C(l)$ vs $l/\lambda_0$ are the same. Taking the Fourier transform of both, we have 
\be
P(f\tau_0)=E(k\lambda_0)
\label{eq:equivalence}
\ee
with $f\tau_0 = k\lambda_0$. Thus, the normalized spectrum $E(k\lambda_0)$ in wave-number space is the same (to second order) as $P(f\tau_0)$ in the frequency domain \cite{HHT10}, and all arguments regarding the dependence of $E(k\lambda_0)$ on $k\lambda_0$ apply equally well to the dependence of $P(f\tau_0)$ on $f\tau_0$. Since our measurements are in the time domain, we shall continue to present our results in the form of $P(f \tau_0)$, but keep in mind that $E(k \lambda_0)$ would have all the same properties.

In the log layer of turbulent shear flows, similarity analysis yielded the prediction \cite{PA77,PHC86} that the wave-number spectrum of the velocity  should be 
proportional to $k^{-1}$ in the intermediate wave-number range, and that this dependence on the wave number implies 
the logarithmic  dependence of the velocity variance on position in space. The arguments leading to these results are quite general and one may expect by analogy that they also apply to the temperature variance. As we shall show below, our data are indeed consistent with $P(f \tau_0) \sim (f \tau_0)^{-1}$ (or equivalently in the spatial domain $E(k \lambda_0) \sim (k \lambda_0)^{-1}$) in the region where we also found that $\sigma^2(z,r)$ was consistent with a logarithmic dependence on $z/L$.

\begin{figure}
  \centerline{\includegraphics[width=0.9 \columnwidth]{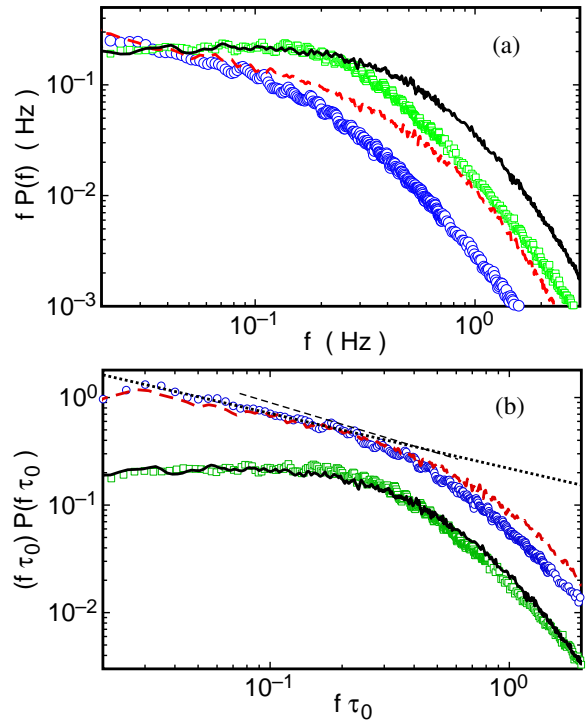}}
  \caption{(color online). (a): Normalized temperature power spectra $P(f)$, compensated by the frequency $f$ in Hz, as a function of $f$. (b): Normalized spectra, compensated by $f \tau_0$, as a function of $f\tau_0$. All data are for the radial location $(R-r)/D = 0.0178$. They are for $z/L=0.4933$, $\Ra=1.08\times10^{15}$ (red dashed curve); $z/L=0.4933$, $\Ra=1.63\times10^{13}$ (blue circles); $z/L=0.0179$, $\Ra=1.08\times10^{15}$ (black solid curve); and $z/L=0.0179$, $\Ra=1.63\times10^{13}$ (green squares). The dashed (dotted) line in (b) corresponds to $P(f\tau_0) \sim (f\tau_0)^{-5/3}$ [$P(f\tau_0) \sim (f\tau_0)^{-1.5}$].}
\label{fig:P_Ra}
\end{figure}

Finally, we turn to experimental results for the spectra. Figure \ref{fig:P_Ra} (a) shows compensated normalized temperature frequency power spectra $fP(f)$ as a function of $f$ from measurements at the radial position $(R-r)/D = 0.0178$ for two vertical positions: one is in the log layer ($z/L=0.0179$) and the other is at the sample midheight ($z/L = 0.493$). At each position measurements are shown for $\Ra = 1.08 \times 10^{15}$ and $1.63 \times 10^{13}$, representing the ultimate and the classical state of RBC respectively. One sees that the shape of the spectrum changes with positions as well as with $\Ra$. 

Using the value of $\tau_0$ derived from the autocorrelation function corresponding to each spectrum (see Table \ref{tab:parameters}), we determined the normalized $P(f\tau_0)$ from the spectra in Fig. \ref{fig:P_Ra} (a) and plotted the compensated spectra $(f\tau_0) P(f\tau_0)$ as a function of $f\tau_0$ in  Fig. \ref{fig:P_Ra} (b). The four spectra now fall into two distinct groups. In the log layer we find a universal spectrum (i.e., independent of \Ra) which, over the range $0.02 \alt f \tau_0 \alt 0.2$, is described well by $P(f\tau_0) = P_0 \times (f\tau_0)^{-\alpha}$ with $\alpha \simeq 1.0$ and $P_0 \simeq 0.208$. This observation suggests that the attached-eddy hypothesis of Ref. \cite{To76} and the similarity argument of Ref. \cite{PHC86} for near-wall turbulence similarly apply to the RBC bulk temperature (a passive scalar \cite{SZX06, HCT11}). Elsewhere \cite {ABH14} we discuss in more detail these similarities for the classical state and suggest that the plumes in RBC may play a role similar to the coherent eddies in shear flow, and that the inner length scale given by the viscous sublayer in shear flows (where the coherent eddies originate) is given by the thermal boundary-layer thickness $\lambda_{th}$ in RBC (where the plumes originate). However, the observed universality persists even though one spectrum is in the classical and the other in the ultimate regime of RBC, and in the ultimate state a different length scale (perhaps that of the thermal sublayer) must prevail. 

At midheight, far above the log layer, the experimental spectrum suggests values of $\alpha$ near 1.5 or larger, as indicated by the dotted ($\alpha = 1.5$) and dashed ($\alpha = 5/3$) lines in Fig. \ref{fig:P_Ra} (b). The value $5/3$ would be consistent with Obukhov-Corrsin scaling for passive scalars \cite{Ob49, Co51}, and suggests correspondence with experimental findings of Kolmogorov scaling along the center line of turbulent pipe flow \cite{RHVBS13}. Temperature power spectra have been studied before  both by experiment \cite{SWL89, PCCKLW91, CCS95, ZX01} and theory \cite{Ca90, Lv91, Lo94} and have yielded values of $\alpha$ ranging from $1.30$ to $1.67$. Our spectrum  at midheight is very similar to  results from previous studies \cite{NSSD00}; like our data, those results could be interpreted to yield an effective $\alpha$ near $7/5$ at low and near $5/3$ at slightly higher frequencies.

\begin{figure}
  \centerline{\includegraphics[width=0.9 \columnwidth]{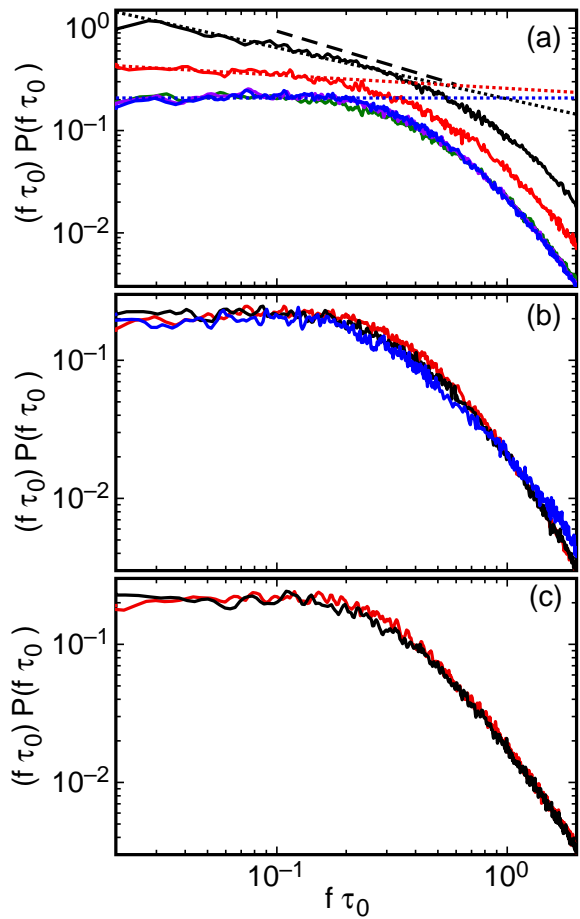}}
  \caption{(color online). Normalized temperature spectra, compensated by $f\tau_0$, as a function of $f\tau_0$. (a): Spectra for radial location $(R-r)/D = 0.0178$ and vertical locations $z/L=0.0179$ (green), $0.0362$ (purple), $0.0719$ (blue), $0.1438$ (red), and $0.4933$ (black). For clarity, the red and black curves were offset by a factor of 2 and 4, respectively. The dashed line represents $P(f\tau_0) \sim (f\tau_0)^{-5/3}$ while the dotted lines are power-law fits to the spectra for $f\tau_0 \leq 0.2$. (b) Spectra for the vertical location $z/L = 0.0719$ and radial locations $(R-r)/D = 0.0178$ (red), $0.0356$ (black), and $0.1337$ (blue). Both (a) and (b) are measured for $\Ra=1.08\times10^{15}$. (c) Spectra for the vertical location $z/L = 0.0179$ and radial locations $(R-r)/D = 0.0178$ (red) and $0.1337$ (black) for $\Ra=1.63\times10^{13}$.
  }
\label{fig:P_z}
\end{figure}

Figure \ref{fig:P_z} (a) shows compensated spectra for $\Ra = 1.08 \times 10^{15}$ measured at the fixed radial position $(R-r)/D = 0.0178$ for varying vertical positions. For the three positions inside the log layer, the spectra again are universal and give $\alpha \simeq 1.0$ for $0.02 \alt f \tau_0 \alt 0.2$. As $z/L$ increases beyond about $0.1$, the spectra depart from the universal form and $\alpha$ increases beyond the value $1.0$ found in the log layer. At the sample midheight, $\alpha$ is in the range of 1.5 (dotted line) to 5/3 (dashed line) as seen before in Fig.~\ref{fig:P_Ra}. For the position between the log layer and the horizontal midplane (red curve), the value of $\alpha$ has an intermediate value, suggesting a gradual transition  from near-wall turbulence to free turbulence. 

Figure \ref{fig:P_z} (b) shows compensated spectra measured at the fixed height $z/L = 0.0719$ for different radial locations and $\Ra = 1.08 \times 10^{15}$. In both Figs. \ref{fig:P_z} (a) and (b), all spectra in the log layer fall on a universal curve independent of position. Figure \ref{fig:P_z} (c) shows compensated spectra measured at heights $z/L = 0.0179$ for different radial locations measured at $\Ra = 1.63 \times 10^{13}$. These spectra also collapse onto the same curve as all the others in the log layer, indicating that $P(f\tau_0) = P_0 \times (f \tau_0)^{-1}$ inside the log layer is universal in the sense that it is independent of both radial and axial positions and of \Ra. Remarkably as shown in Table~\ref{tab:parameters}, a universal value $P_0 = 0.208 \pm 0.008$ applies for both the ultimate and the classical state of RBC, if $\alpha = 1$ is assumed.  

In this Letter we presented the results of a systematic experimental study of the temperature variance $\sigma^2(z,r)$ and of the spectra $P(f,z,r)$ in a $\Gamma = 0.50$ RBC sample for the ultimate state at $\Ra = 1.1\times 10^{15}$ and the classical state at $\Ra = 1.6\times 10^{13}$. For both states we found that $\sigma^2(z,r)$ varied logarithmically with $z/L$ near the bottom plate where $z/L \alt 0.1$. The log layer for $\sigma^2$ shares many similarities with the log layer observed in turbulent shear flows such as pipe flow, channel flow, and Taylor-Couette flow \cite{WW89, HVBS12, HSCKLS13} (regarding this issue, see also \cite{ABH14}). 

On the basis of the elliptic approximation of correlation functions \cite{HZ06} we discussed that there is an equivalence between frequency spectra $P(f \tau_0)$ and wave-number spectra $E(k  \lambda_0)$ when frequency is scaled  by a time scale $\tau_0$ based on the curvature of the time autocorrelation function of the temperature and the wave number is scaled by a similarly defined length scale $\lambda_0$. Thus, predictions for $E(k \lambda_0)$ apply equally well to $P(f \tau_0)$. 

For both \Ra\ values, and for all radial and axial positions where the logarithmic dependence of $\sigma^2$ on $z/L$ was found, we found that $P(f\tau_0) = P_0 \times (f \tau_0)^{-1}$ in the range $0.02 \alt f \tau_0 \alt 0.2$, with $P_0$ a universal constant equal to $0.208 \pm 0.008$. This result, although with its own independent theoretical explanation still missing, shares many similarities with predictions for the wave-number spectrum of velocity fluctuations in the log layer of turbulent pipe flow \cite{PA77,PHC86} where the expected $k^{-1}$ dependence remains ``elusive" \cite{RHVBS13}.

\vskip 0.05in
\noindent {\it Acknowledgements:}

We are grateful to the Max-Planck-Society and the Volkswagen Stiftung, whose generous support made the establishment of the facility and the experiments possible. We thank the Deutsche Forschungsgemeinschaft (DFG) for financial support through SFB963: ``Astrophysical Flow Instabilities and Turbulence."  The work of G.A. was supported in part by the U.S National Science Foundation through Grant No. DMR11-58514. We thank Andreas Kopp, Artur Kubitzek, Holger Nobach, and Andreas Renner for their enthusiastic technical support. 
\vskip -0.2in


\begin{thebibliography}{48}%
\makeatletter
\providecommand \@ifxundefined [1]{%
 \@ifx{#1\undefined}
}%
\providecommand \@ifnum [1]{%
 \ifnum #1\expandafter \@firstoftwo
 \else \expandafter \@secondoftwo
 \fi
}%
\providecommand \@ifx [1]{%
 \ifx #1\expandafter \@firstoftwo
 \else \expandafter \@secondoftwo
 \fi
}%
\providecommand \natexlab [1]{#1}%
\providecommand \enquote  [1]{``#1''}%
\providecommand \bibnamefont  [1]{#1}%
\providecommand \bibfnamefont [1]{#1}%
\providecommand \citenamefont [1]{#1}%
\providecommand \href@noop [0]{\@secondoftwo}%
\providecommand \href [0]{\begingroup \@sanitize@url \@href}%
\providecommand \@href[1]{\@@startlink{#1}\@@href}%
\providecommand \@@href[1]{\endgroup#1\@@endlink}%
\providecommand \@sanitize@url [0]{\catcode `\\12\catcode `\$12\catcode
  `\&12\catcode `\#12\catcode `\^12\catcode `\_12\catcode `\%12\relax}%
\providecommand \@@startlink[1]{}%
\providecommand \@@endlink[0]{}%
\providecommand \url  [0]{\begingroup\@sanitize@url \@url }%
\providecommand \@url [1]{\endgroup\@href {#1}{\urlprefix }}%
\providecommand \urlprefix  [0]{URL }%
\providecommand \Eprint [0]{\href }%
\providecommand \doibase [0]{http://dx.doi.org/}%
\providecommand \selectlanguage [0]{\@gobble}%
\providecommand \bibinfo  [0]{\@secondoftwo}%
\providecommand \bibfield  [0]{\@secondoftwo}%
\providecommand \translation [1]{[#1]}%
\providecommand \BibitemOpen [0]{}%
\providecommand \bibitemStop [0]{}%
\providecommand \bibitemNoStop [0]{.\EOS\space}%
\providecommand \EOS [0]{\spacefactor3000\relax}%
\providecommand \BibitemShut  [1]{\csname bibitem#1\endcsname}%
\let\auto@bib@innerbib\@empty
\bibitem [{\citenamefont {van Doorn}\ \emph {et~al.}(2000)\citenamefont {van
  Doorn}, \citenamefont {Dhruva}, \citenamefont {Sreenivasan},\ and\
  \citenamefont {Cassella}}]{DDSC00}%
  \BibitemOpen
  \bibfield  {author} {\bibinfo {author} {\bibfnamefont {E.}~\bibnamefont {van
  Doorn}}, \bibinfo {author} {\bibfnamefont {B.}~\bibnamefont {Dhruva}},
  \bibinfo {author} {\bibfnamefont {K.~R.}\ \bibnamefont {Sreenivasan}}, \ and\
  \bibinfo {author} {\bibfnamefont {V.}~\bibnamefont {Cassella}},\ }\href@noop
  {} {\bibfield  {journal} {\bibinfo  {journal} {Phys. Fluids}\ }\textbf
  {\bibinfo {volume} {12}},\ \bibinfo {pages} {1529} (\bibinfo {year}
  {2000})}\BibitemShut {NoStop}%
\bibitem [{\citenamefont {Marshall}\ and\ \citenamefont {Schott}(1999)}]{MS99}%
  \BibitemOpen
  \bibfield  {author} {\bibinfo {author} {\bibfnamefont {J.}~\bibnamefont
  {Marshall}}\ and\ \bibinfo {author} {\bibfnamefont {F.}~\bibnamefont
  {Schott}},\ }\href@noop {} {\bibfield  {journal} {\bibinfo  {journal} {Rev.
  Geophys.}\ }\textbf {\bibinfo {volume} {37}},\ \bibinfo {pages} {1} (\bibinfo
  {year} {1999})}\BibitemShut {NoStop}%
\bibitem [{\citenamefont {Cardin}\ and\ \citenamefont {Olson}(1994)}]{CO94}%
  \BibitemOpen
  \bibfield  {author} {\bibinfo {author} {\bibfnamefont {P.}~\bibnamefont
  {Cardin}}\ and\ \bibinfo {author} {\bibfnamefont {P.}~\bibnamefont {Olson}},\
  }\href@noop {} {\bibfield  {journal} {\bibinfo  {journal} {Phys. Earth Planet. Inter.}\ }\textbf {\bibinfo {volume} {82}},\ \bibinfo
  {pages} {235} (\bibinfo {year} {1994})}\BibitemShut {NoStop}%
\bibitem [{\citenamefont {Busse}(1994)}]{Bu94}%
  \BibitemOpen
  \bibfield  {author} {\bibinfo {author} {\bibfnamefont {F.~H.}\ \bibnamefont
  {Busse}},\ }\href@noop {} {\bibfield  {journal} {\bibinfo  {journal} {Chaos}\
  }\textbf {\bibinfo {volume} {4}},\ \bibinfo {pages} {123} (\bibinfo {year}
  {1994})}\BibitemShut {NoStop}%
\bibitem [{\citenamefont {Ahlers}(2009)}]{Ah09}%
  \BibitemOpen
  \bibfield  {author} {\bibinfo {author} {\bibfnamefont {G.}~\bibnamefont
  {Ahlers}},\ }\href@noop {} {\bibfield  {journal} {\bibinfo  {journal}
  {Physics}\ }\textbf {\bibinfo {volume} {2}},\ \bibinfo {pages} {74} (\bibinfo
  {year} {2009})}\BibitemShut {NoStop}%
\bibitem [{\citenamefont {Ahlers}\ \emph
  {et~al.}(2009{\natexlab{a}})\citenamefont {Ahlers}, \citenamefont
  {Grossmann},\ and\ \citenamefont {Lohse}}]{AGL09}%
  \BibitemOpen
  \bibfield  {author} {\bibinfo {author} {\bibfnamefont {G.}~\bibnamefont
  {Ahlers}}, \bibinfo {author} {\bibfnamefont {S.}~\bibnamefont {Grossmann}}, \
  and\ \bibinfo {author} {\bibfnamefont {D.}~\bibnamefont {Lohse}},\
  }\href@noop {} {\bibfield  {journal} {\bibinfo  {journal} {Rev. Mod. Phys.}\
  }\textbf {\bibinfo {volume} {81}},\ \bibinfo {pages} {503} (\bibinfo {year}
  {2009}{\natexlab{a}})}\BibitemShut {NoStop}%
\bibitem [{\citenamefont {Lohse}\ and\ \citenamefont {Xia}(2010)}]{LX10}%
  \BibitemOpen
  \bibfield  {author} {\bibinfo {author} {\bibfnamefont {D.}~\bibnamefont
  {Lohse}}\ and\ \bibinfo {author} {\bibfnamefont {K.-Q.}\ \bibnamefont
  {Xia}},\ }\href@noop {} {\bibfield  {journal} {\bibinfo  {journal} {Annu.
  Rev. Fluid Mech.}\ }\textbf {\bibinfo {volume} {42}},\ \bibinfo {pages} {335}
  (\bibinfo {year} {2010})}\BibitemShut {NoStop}%
\bibitem [{\citenamefont {Chill\'a}\ and\ \citenamefont
  {Schumacher}(2012)}]{CS12}%
  \BibitemOpen
  \bibfield  {author} {\bibinfo {author} {\bibfnamefont {F.}~\bibnamefont
  {Chill\'a}}\ and\ \bibinfo {author} {\bibfnamefont {J.}~\bibnamefont
  {Schumacher}},\ }\href@noop {} {\bibfield  {journal} {\bibinfo  {journal}
  {Eur. Phys. J. E}\ }\textbf {\bibinfo {volume} {35}},\ \bibinfo {pages} {58}
  (\bibinfo {year} {2012})}\BibitemShut {NoStop}%
\bibitem [{\citenamefont {Kraichnan}(1962)}]{Kr62}%
  \BibitemOpen
  \bibfield  {author} {\bibinfo {author} {\bibfnamefont {R.~H.}\ \bibnamefont
  {Kraichnan}},\ }\href@noop {} {\bibfield  {journal} {\bibinfo  {journal}
  {Phys. Fluids}\ }\textbf {\bibinfo {volume} {5}},\ \bibinfo {pages} {1374}
  (\bibinfo {year} {1962})}\BibitemShut {NoStop}%
\bibitem [{\citenamefont {Spiegel}(1971)}]{Sp71}%
  \BibitemOpen
  \bibfield  {author} {\bibinfo {author} {\bibfnamefont {E.~A.}\ \bibnamefont
  {Spiegel}},\ }\href@noop {} {\bibfield  {journal} {\bibinfo  {journal} {Ann.
  Rev. Astron. Astrophys.}\ }\textbf {\bibinfo {volume} {9}},\ \bibinfo {pages}
  {323} (\bibinfo {year} {1971})}\BibitemShut {NoStop}%
\bibitem [{\citenamefont {Grossmann}\ and\ \citenamefont {Lohse}(2011)}]{GL11}%
  \BibitemOpen
  \bibfield  {author} {\bibinfo {author} {\bibfnamefont {S.}~\bibnamefont
  {Grossmann}}\ and\ \bibinfo {author} {\bibfnamefont {D.}~\bibnamefont
  {Lohse}},\ }\href@noop {} {\bibfield  {journal} {\bibinfo  {journal} {Phys.
  Fluids}\ }\textbf {\bibinfo {volume} {23}},\ \bibinfo {pages} {045108}
  (\bibinfo {year} {2011})}\BibitemShut {NoStop}%
\bibitem [{\citenamefont {He}\ \emph {et~al.}(2012)\citenamefont {He},
  \citenamefont {Funfschilling}, \citenamefont {Nobach}, \citenamefont
  {Bodenschatz},\ and\ \citenamefont {Ahlers}}]{HFNBA12}%
  \BibitemOpen
  \bibfield  {author} {\bibinfo {author} {\bibfnamefont {X.}~\bibnamefont
  {He}}, \bibinfo {author} {\bibfnamefont {D.}~\bibnamefont {Funfschilling}},
  \bibinfo {author} {\bibfnamefont {H.}~\bibnamefont {Nobach}}, \bibinfo
  {author} {\bibfnamefont {E.}~\bibnamefont {Bodenschatz}}, \ and\ \bibinfo
  {author} {\bibfnamefont {G.}~\bibnamefont {Ahlers}},\ }\href@noop {}
  {\bibfield  {journal} {\bibinfo  {journal} {Phys. Rev. Lett.}\ }\textbf
  {\bibinfo {volume} {108}},\ \bibinfo {pages} {024502} (\bibinfo {year}
  {2012})}\BibitemShut {NoStop}%
\bibitem [{\citenamefont {Ahlers}\ \emph
  {et~al.}(2012{\natexlab{a}})\citenamefont {Ahlers}, \citenamefont
  {Bodenschatz}, \citenamefont {Funfschilling}, \citenamefont {Grossmann},
  \citenamefont {He}, \citenamefont {Lohse}, \citenamefont {Stevens},\ and\
  \citenamefont {Verzicco}}]{ABFGHLSV12}%
  \BibitemOpen
  \bibfield  {author} {\bibinfo {author} {\bibfnamefont {G.}~\bibnamefont
  {Ahlers}}, \bibinfo {author} {\bibfnamefont {E.}~\bibnamefont {Bodenschatz}},
  \bibinfo {author} {\bibfnamefont {D.}~\bibnamefont {Funfschilling}}, \bibinfo
  {author} {\bibfnamefont {S.}~\bibnamefont {Grossmann}}, \bibinfo {author}
  {\bibfnamefont {X.}~\bibnamefont {He}}, \bibinfo {author} {\bibfnamefont
  {D.}~\bibnamefont {Lohse}}, \bibinfo {author} {\bibfnamefont {R.~J. A.~M.}\
  \bibnamefont {Stevens}}, \ and\ \bibinfo {author} {\bibfnamefont
  {R.}~\bibnamefont {Verzicco}},\ }\href@noop {} {\bibfield  {journal}
  {\bibinfo  {journal} {Phys. Rev. Lett.}\ }\textbf {\bibinfo {volume} {109}},\
  \bibinfo {pages} {114501} (\bibinfo {year} {2012}{\natexlab{a}})}\BibitemShut
  {NoStop}%
\bibitem [{\citenamefont {Ahlers}\ \emph {et~al.}(2014)\citenamefont {Ahlers},
  \citenamefont {Bodenschatz},\ and\ \citenamefont {He}}]{ABH14}%
  \BibitemOpen
  \bibfield  {author} {\bibinfo {author} {\bibfnamefont {G.}~\bibnamefont
  {Ahlers}}, \bibinfo {author} {\bibfnamefont {E.}~\bibnamefont {Bodenschatz}},
  \ and\ \bibinfo {author} {\bibfnamefont {X.}~\bibnamefont {He}},\ }\href@noop
  {} {\bibfield  {journal} {\bibinfo  {journal} {arXiv:1404.3459}\
  }}\BibitemShut {NoStop}%
\bibitem [{\citenamefont {Wei}\ and\ \citenamefont {Willmarth}(1989)}]{WW89}%
  \BibitemOpen
  \bibfield  {author} {\bibinfo {author} {\bibfnamefont {T.}~\bibnamefont
  {Wei}}\ and\ \bibinfo {author} {\bibfnamefont {W.~W.}\ \bibnamefont
  {Willmarth}},\ }\href@noop {} {\bibfield  {journal} {\bibinfo  {journal} {J.
  Fluid Mech.}\ }\textbf {\bibinfo {volume} {204}},\ \bibinfo {pages} {57}
  (\bibinfo {year} {1989})}\BibitemShut {NoStop}%
\bibitem [{\citenamefont {Hultmark}\ \emph {et~al.}(2012)\citenamefont
  {Hultmark}, \citenamefont {Vallikivi}, \citenamefont {Bailey},\ and\
  \citenamefont {Smits}}]{HVBS12}%
  \BibitemOpen
  \bibfield  {author} {\bibinfo {author} {\bibfnamefont {M.}~\bibnamefont
  {Hultmark}}, \bibinfo {author} {\bibfnamefont {M.}~\bibnamefont {Vallikivi}},
  \bibinfo {author} {\bibfnamefont {S.~C.~C.}\ \bibnamefont {Bailey}}, \ and\
  \bibinfo {author} {\bibfnamefont {A.~J.}\ \bibnamefont {Smits}},\ }\href@noop
  {} {\bibfield  {journal} {\bibinfo  {journal} {Phys. Rev. Lett.}\ }\textbf
  {\bibinfo {volume} {108}},\ \bibinfo {pages} {094501} (\bibinfo {year}
  {2012})}\BibitemShut {NoStop}%
\bibitem [{\citenamefont {Huisman}\ \emph {et~al.}(2013)\citenamefont
  {Huisman}, \citenamefont {Scharnowski}, \citenamefont {Cierpka},
  \citenamefont {K\"ahler}, \citenamefont {Lohse},\ and\ \citenamefont
  {Sun}}]{HSCKLS13}%
  \BibitemOpen
  \bibfield  {author} {\bibinfo {author} {\bibfnamefont {S.~G.}\ \bibnamefont
  {Huisman}}, \bibinfo {author} {\bibfnamefont {S.}~\bibnamefont
  {Scharnowski}}, \bibinfo {author} {\bibfnamefont {C.}~\bibnamefont
  {Cierpka}}, \bibinfo {author} {\bibfnamefont {C.~J.}\ \bibnamefont
  {K\"ahler}}, \bibinfo {author} {\bibfnamefont {D.}~\bibnamefont {Lohse}}, \
  and\ \bibinfo {author} {\bibfnamefont {C.}~\bibnamefont {Sun}},\ }\href@noop
  {} {\bibfield  {journal} {\bibinfo  {journal} {Phys. Rev. Lett.}\ }\textbf
  {\bibinfo {volume} {110}},\ \bibinfo {pages} {264501} (\bibinfo {year}
  {2013})}\BibitemShut {NoStop}%
\bibitem [{\citenamefont {Prandtl}(1925)}]{Pr25}%
  \BibitemOpen
  \bibfield  {author} {\bibinfo {author} {\bibfnamefont {L.}~\bibnamefont
  {Prandtl}},\ }\href@noop {} {\bibfield  {journal} {\bibinfo  {journal} {Z.
  Angew. Math. Mech.}\ }\textbf {\bibinfo {volume} {5}},\ \bibinfo {pages}
  {136} (\bibinfo {year} {1925})}\BibitemShut {NoStop}%
\bibitem [{\citenamefont {{von K\'arm\'an}}(1930)}]{Ka30}%
  \BibitemOpen
  \bibfield  {author} {\bibinfo {author} {\bibfnamefont {T.}~\bibnamefont {{von
  K\'arm\'an}}},\ }\href@noop {} {\bibfield  {journal} {\bibinfo  {journal}
  {Nachr. Ges. Wiss. {G\"ottingen}, Math.-Phys. Kl.}\ }{\bibinfo
  {pages} {58}} (\bibinfo {year} {1930})}\BibitemShut {NoStop}%
\bibitem [{\citenamefont {Prandtl}(1932)}]{Pr32}%
  \BibitemOpen
  \bibfield  {author} {\bibinfo {author} {\bibfnamefont {L.}~\bibnamefont
  {Prandtl}},\ }\href@noop {} {\bibfield  {journal} {\bibinfo  {journal}
  {Ergeb. Aerodyn. Versuch, {G\"ottingen}}\ \textbf {\bibinfo {volume} {4}},\ \bibinfo {pages} {18}} (\bibinfo
  {year} {1932})}\BibitemShut {NoStop}%
\bibitem [{\citenamefont {Marusic}\ \emph {et~al.}(2010)\citenamefont
  {Marusic}, \citenamefont {McKeon}, \citenamefont {Monkewitz}, \citenamefont
  {Nagib}, \citenamefont {Smits},\ and\ \citenamefont
  {Sreenivasan}}]{MMMNSS10}%
  \BibitemOpen
  \bibfield  {author} {\bibinfo {author} {\bibfnamefont {I.}~\bibnamefont
  {Marusic}}, \bibinfo {author} {\bibfnamefont {B.~J.}\ \bibnamefont {McKeon}},
  \bibinfo {author} {\bibfnamefont {P.~A.}\ \bibnamefont {Monkewitz}}, \bibinfo
  {author} {\bibfnamefont {H.~M.}\ \bibnamefont {Nagib}}, \bibinfo {author}
  {\bibfnamefont {A.~J.}\ \bibnamefont {Smits}}, \ and\ \bibinfo {author}
  {\bibfnamefont {K.~R.}\ \bibnamefont {Sreenivasan}},\ }\href@noop {}
  {\bibfield  {journal} {\bibinfo  {journal} {Phys. Fluids}\ }\textbf {\bibinfo
  {volume} {22}},\ \bibinfo {pages} {065103} (\bibinfo {year}
  {2010})}\BibitemShut {NoStop}%
\bibitem [{\citenamefont {Smits}\ \emph {et~al.}(2011)\citenamefont
  {Smits}, \citenamefont {McKeon},\ and\ \citenamefont {Marusic}}]{SMM11}%
  \BibitemOpen
  \bibfield  {author} {\bibinfo{author} {\bibfnamefont {A.~J.}\ \bibnamefont {Smits}}, \bibinfo {author}
  {\bibfnamefont {B.~J.}\ \bibnamefont {McKeon}},\ and\ \bibinfo {author}
  {\bibfnamefont {Ivan}\ \bibnamefont {Marusic}},\ }\href@noop {}
  {\bibfield  {journal} {\bibinfo  {journal} {Annu. Rev. Fluid Mech.}\ }\textbf {\bibinfo
  {volume} {43}},\ \bibinfo {pages} {353} (\bibinfo {year}
  {2011})}\BibitemShut {NoStop}%
\bibitem [{\citenamefont {Grossmann}\ and\ \citenamefont {Lohse}(2012)}]{GL12}%
  \BibitemOpen
  \bibfield  {author} {\bibinfo {author} {\bibfnamefont {S.}~\bibnamefont
  {Grossmann}}\ and\ \bibinfo {author} {\bibfnamefont {D.}~\bibnamefont
  {Lohse}},\ }\href@noop {} {\bibfield  {journal} {\bibinfo  {journal} {Phys.
  Fluids}\ }\textbf {\bibinfo {volume} {24}},\ \bibinfo {pages} {125103}
  (\bibinfo {year} {2012})}\BibitemShut {NoStop}%
\bibitem [{\citenamefont {She}\ \emph {et~al.}(2014)\citenamefont {She},
  \citenamefont {Chen},\citenamefont {Chen},\citenamefont {Zou},\citenamefont {Bao},\ and\ \citenamefont {Hussain}}]{SCCZBH14}%
  \BibitemOpen
  \bibfield  {author} {\bibinfo {author} {\bibfnamefont {Z.-S.}~\bibnamefont
  {She}}, \bibinfo {author} {\bibfnamefont {X.}~\bibnamefont
  {Chen}}, \bibinfo {author} {\bibfnamefont {J.}~\bibnamefont
  {Chen}}, \bibinfo {author} {\bibfnamefont {H.-Y.}~\bibnamefont
  {Zou}}, \bibinfo {author} {\bibfnamefont {Y.}~\bibnamefont {Bao}},
  \ and\ \bibinfo {author} {\bibfnamefont {F.}~\bibnamefont {Hussain}},\ }\href@noop
  {} {\bibfield  {journal} {\bibinfo  {journal} {arXiv:1401.2138v1}\
  }}\BibitemShut {NoStop}%
\bibitem [{\citenamefont {Obukhov}(1949)}]{Ob49}%
  \BibitemOpen
  \bibfield  {author} {\bibinfo {author} {\bibfnamefont {A.~M.}\ \bibnamefont
  {Obukhov}},\ }\href@noop {} {\bibfield  {journal} {\bibinfo  {journal} {Izv.
  Akad. Nauk SSSR, Ser. Geog. Geofiz.}\ }\textbf {\bibinfo {volume} {13}},\
  \bibinfo {pages} {58} (\bibinfo {year} {1949})}\BibitemShut {NoStop}%
\bibitem [{\citenamefont {Corrsin}(1951)}]{Co51}%
  \BibitemOpen
  \bibfield  {author} {\bibinfo {author} {\bibfnamefont {S.}~\bibnamefont
  {Corrsin}},\ }\href@noop {} {\bibfield  {journal} {\bibinfo  {journal} {J.
  Appl. Phys.}\ }\textbf {\bibinfo {volume} {22}},\ \bibinfo {pages} {469}
  (\bibinfo {year} {1951})}\BibitemShut {NoStop}%
\bibitem [{\citenamefont {He}\ and\ \citenamefont {Zhang}(2006)}]{HZ06}%
  \BibitemOpen
  \bibfield  {author} {\bibinfo {author} {\bibfnamefont {G.-W}~\bibnamefont
  {He}}\ and\ \bibinfo {author} {\bibfnamefont {J.-B.}~\bibnamefont
  {Zhang}},\ }\href@noop {} {\bibfield  {journal} {\bibinfo  {journal} {Phys. Rev. E}\ }\textbf {\bibinfo {volume} {73}},\ \bibinfo {pages} {055303}
  (\bibinfo {year} {2006})}\BibitemShut {NoStop}%
\bibitem [{\citenamefont {Ahlers}\ \emph
  {et~al.}(2009{\natexlab{b}})\citenamefont {Ahlers}, \citenamefont
  {Funfschilling},\ and\ \citenamefont {Bodenschatz}}]{AFB09}%
  \BibitemOpen
  \bibfield  {author} {\bibinfo {author} {\bibfnamefont {G.}~\bibnamefont
  {Ahlers}}, \bibinfo {author} {\bibfnamefont {D.}~\bibnamefont
  {Funfschilling}}, \ and\ \bibinfo {author} {\bibfnamefont {E.}~\bibnamefont
  {Bodenschatz}},\ }\href@noop {} {\bibfield  {journal} {\bibinfo  {journal}
  {New J. Phys.}\ }\textbf {\bibinfo {volume} {11}},\ \bibinfo {pages} {123001}
  (\bibinfo {year} {2009}{\natexlab{b}})}\BibitemShut {NoStop}%
\bibitem [{\citenamefont {Ahlers}\ \emph
  {et~al.}(2012{\natexlab{b}})\citenamefont {Ahlers}, \citenamefont {He},
  \citenamefont {Funfschilling},\ and\ \citenamefont {Bodenschatz}}]{AHFB12}%
  \BibitemOpen
  \bibfield  {author} {\bibinfo {author} {\bibfnamefont {G.}~\bibnamefont
  {Ahlers}}, \bibinfo {author} {\bibfnamefont {X.}~\bibnamefont {He}}, \bibinfo
  {author} {\bibfnamefont {D.}~\bibnamefont {Funfschilling}}, \ and\ \bibinfo
  {author} {\bibfnamefont {E.}~\bibnamefont {Bodenschatz}},\ }\href@noop {}
  {\bibfield  {journal} {\bibinfo  {journal} {New J. Phys.}\ }\textbf {\bibinfo
  {volume} {14}},\ \bibinfo {pages} {103012} (\bibinfo {year}
  {2012}{\natexlab{b}})}\BibitemShut {NoStop}%
\bibitem [{\citenamefont {He}\ \emph {et~al.}(2013)\citenamefont {He},
  \citenamefont {Funfschilling}, \citenamefont {Nobach}, \citenamefont
  {Bodenschatz},\ and\ \citenamefont {Ahlers}}]{HFNBA13}%
  \BibitemOpen
  \bibfield  {author} {\bibinfo {author} {\bibfnamefont {X.}~\bibnamefont
  {He}}, \bibinfo {author} {\bibfnamefont {D.}~\bibnamefont {Funfschilling}},
  \bibinfo {author} {\bibfnamefont {H.}~\bibnamefont {Nobach}}, \bibinfo
  {author} {\bibfnamefont {E.}~\bibnamefont {Bodenschatz}}, \ and\ \bibinfo
  {author} {\bibfnamefont {G.}~\bibnamefont {Ahlers}},\ }\href@noop {}
  {\bibfield  {journal} {\bibinfo  {journal} {Phys. Rev. Lett.}\ }\textbf
  {\bibinfo {volume} {110}},\ \bibinfo {pages} {199401} (\bibinfo {year}
  {2013})}\BibitemShut {NoStop}%
  \bibitem [{\citenamefont {He}\ \emph {et~al.}(2013)\citenamefont {He},
  \citenamefont {Funfschilling}, \citenamefont {Nobach}, \citenamefont
  {Bodenschatz},\ and\ \citenamefont {Ahlers}}]{SD}%
  \BibitemOpen
 {{\bibinfo  {journal} {See Supplemental Material for details of the data acquisition, the data analysis, important aspects of the elliptic approximation, and of other discussions relevant to the present paper, which includes Ref. \cite{ZH09,Ko41,HT11,ZLLL11}}}}
\BibitemShut {NoStop}%
\bibitem [{\citenamefont {Zhao}\ and\ \citenamefont
  {He}(2009)}]{ZH09}%
  \BibitemOpen
  \bibfield  {author} {\bibinfo {author} {\bibfnamefont {X.}~\bibnamefont
  {Zhao}}\ and\ \bibinfo {author} {\bibfnamefont {G.-W.}~\bibnamefont
  {He}},\ }\href@noop {} {\bibfield  {journal} {\bibinfo  {journal}
  {Phys. Rev. E}\ }\textbf {\bibinfo {volume} {79}},\ \bibinfo {pages} {046316}
  (\bibinfo {year} {2009})}\BibitemShut {NoStop}%
\bibitem [{\citenamefont {Kolmogorov}(1941)}]{Ko41}%
  \BibitemOpen
  \bibfield  {author} {\bibinfo {author} {\bibfnamefont {A.~N.}\ \bibnamefont
  {Kolmogorov}},\ }\href@noop {} {\bibfield  {journal} {\bibinfo  {journal}
  {CR. Acad. Sci. USSR}\ }\textbf {\bibinfo {volume} {30}},\ \bibinfo {pages} {299}
  (\bibinfo {year} {1941})}\BibitemShut {NoStop}%
\bibitem [{\citenamefont {He}\ and\ \citenamefont
  {Tong}(2011)}]{HT11}%
  \BibitemOpen
  \bibfield  {author} {\bibinfo {author} {\bibfnamefont {X.}~\bibnamefont
  {He}}\ and\ \bibinfo {author} {\bibfnamefont {P.}~\bibnamefont
  {Tong}},\ }\href@noop {} {\bibfield  {journal} {\bibinfo  {journal}
  {Phys. Rev. E}\ }\textbf {\bibinfo {volume} {83}},\ \bibinfo {pages} {037302}
  (\bibinfo {year} {2011})}\BibitemShut {NoStop}%
\bibitem [{\citenamefont {Zhou}\ \emph
  {et~al.}(2011)\citenamefont {Zhou}, \citenamefont {Li},
  \citenamefont {Lu},\ and\ \citenamefont {Liu}}]{ZLLL11}%
  \BibitemOpen
  \bibfield  {author} {\bibinfo {author} {\bibfnamefont {Q.}~\bibnamefont
  {Zhou}}, \bibinfo {author} {\bibfnamefont {C.-M.}~\bibnamefont {Li}}, \bibinfo
  {author} {\bibfnamefont {Z.-M.}~\bibnamefont {Lu}}, \ and\ \bibinfo
  {author} {\bibfnamefont {Y.-L.}~\bibnamefont {Liu}},\ }\href@noop {}
  {\bibfield  {journal} {\bibinfo  {journal} {J. Fluid Mech.}\ }\textbf {\bibinfo
  {volume} {683}},\ \bibinfo {pages} {94} (\bibinfo {year}
  {2011}{\natexlab{b}})}\BibitemShut {NoStop}%
\bibitem [{\citenamefont {Pope}(2000)}]{Po00}%
 \BibitemOpen
  \bibfield  {author} {\bibinfo {author} {\bibfnamefont {S.~B.}\ \bibnamefont
  {Pope}},} {\it{\bibinfo  {journal} {Turbulent Flows}}} {\bibinfo  {journal} {(Cambridge University Press, Cambridge, England, 2000)}}\BibitemShut {NoStop}%
\bibitem [{\citenamefont {He}\ \emph {et~al.}(2013)\citenamefont {He},
  \citenamefont {van Gils}, \citenamefont {Bodenschatz},\ and\ \citenamefont
  {Ahlers}}]{HGBA13b}%
  \BibitemOpen
  \bibfield  {author} {\bibinfo {author} {\bibfnamefont {X.}~\bibnamefont
  {He}}, \bibinfo {author} {\bibfnamefont {D.~P.~M.}\ \bibnamefont {van Gils}},
  \bibinfo {author} {\bibfnamefont {E.}~\bibnamefont {Bodenschatz}}, \ and\
  \bibinfo {author} {\bibfnamefont {G.}~\bibnamefont {Ahlers}}\ }\href@noop {}
  {\bibfield  {journal} {\bibinfo  {journal} {(to be published)}\ }}\BibitemShut {NoStop}%
\bibitem [{\citenamefont {He}\ \emph {et~al.}(2010)\citenamefont {He},
  \citenamefont {He},\ and\ \citenamefont {Tong}}]{HHT10}%
  \BibitemOpen
  \bibfield  {author} {\bibinfo {author} {\bibfnamefont {X.}~\bibnamefont
  {He}}, \bibinfo {author} {\bibfnamefont {G.}~\bibnamefont {He}}, \ and\
  \bibinfo {author} {\bibfnamefont {P.}~\bibnamefont {Tong}},\ }\href@noop {}
  {\bibfield  {journal} {\bibinfo  {journal} {Phys. Rev. E}\ }\textbf {\bibinfo
  {volume} {81}},\ \bibinfo {pages} {065303} (\bibinfo {year}
  {2010})}\BibitemShut {NoStop}%
\bibitem [{\citenamefont {Perry}\ and\ \citenamefont {Abell}(1977)}]{PA77}%
  \BibitemOpen
  \bibfield  {author} {\bibinfo {author} {\bibfnamefont {A.~E.}\ \bibnamefont
  {Perry}}\ and\ \bibinfo {author} {\bibfnamefont {C.~J.}\ \bibnamefont
  {Abell}},\ }\href@noop {} {\bibfield  {journal} {\bibinfo  {journal} {J.
  Fluid Mech.}\ }\textbf {\bibinfo {volume} {79}},\ \bibinfo {pages} {785}
  (\bibinfo {year} {1977})}\BibitemShut {NoStop}%
\bibitem [{\citenamefont {Perry}\ \emph {et~al.}(1986)\citenamefont {Perry},
  \citenamefont {Henbest},\ and\ \citenamefont {Chong}}]{PHC86}%
  \BibitemOpen
  \bibfield  {author} {\bibinfo {author} {\bibfnamefont {A.~E.}\ \bibnamefont
  {Perry}}, \bibinfo {author} {\bibfnamefont {S.}~\bibnamefont {Henbest}}, \
  and\ \bibinfo {author} {\bibfnamefont {M.}~\bibnamefont {Chong}},\
  }\href@noop {} {\bibfield  {journal} {\bibinfo  {journal} {J. Fluid Mech.}\
  }\textbf {\bibinfo {volume} {165}},\ \bibinfo {pages} {163} (\bibinfo {year}
  {1986})}\BibitemShut {NoStop}%
\bibitem [{\citenamefont {Townsend}(1976)}]{To76}%
 \BibitemOpen
  \bibfield  {author} {\bibinfo {author} {\bibfnamefont {A.~A.}\ \bibnamefont
  {Townsend}},} {\it{\bibinfo  {journal} {The Structure of
  Turbulent Shear Flow (2nd)}}} {\bibinfo  {journal} {(Cambridge
  University Press, Cambridge, England, 1976)}}\BibitemShut {NoStop}%
  \bibitem [{\citenamefont {Sun}\ \emph {et~al.}(2006)\citenamefont {Sun},
  \citenamefont {Zhou},\ and\ \citenamefont {Xia}}]{SZX06}%
  \BibitemOpen
  \bibfield  {author} {\bibinfo {author} {\bibfnamefont {C.}~\bibnamefont
  {Sun}}, \bibinfo {author} {\bibfnamefont {Q.}~\bibnamefont {Zhou}}, \ and\
  \bibinfo {author} {\bibfnamefont {K.~Q.}\ \bibnamefont {Xia}},\ }\href@noop
  {} {\bibfield  {journal} {\bibinfo  {journal} {Phys. Rev. Lett.}\ }\textbf
  {\bibinfo {volume} {97}},\ \bibinfo {pages} {144504} (\bibinfo {year}
  {2006})}\BibitemShut {NoStop}%
\bibitem [{\citenamefont {He}\ \emph {et~al.}(2011)\citenamefont {He},
  \citenamefont {Ching},\ and\ \citenamefont {Tong}}]{HCT11}%
  \BibitemOpen
  \bibfield  {author} {\bibinfo {author} {\bibfnamefont {X.}~\bibnamefont
  {He}}, \bibinfo {author} {\bibfnamefont {E.~S.~C.}\ \bibnamefont {Ching}}, \
  and\ \bibinfo {author} {\bibfnamefont {P.}~\bibnamefont {Tong}},\ }\href@noop
  {} {\bibfield  {journal} {\bibinfo  {journal} {Phys. of Fluids}\ }\textbf
  {\bibinfo {volume} {23}},\ \bibinfo {pages} {025106} (\bibinfo {year}
  {2011})}\BibitemShut {NoStop}%
\bibitem [{\citenamefont {Rosenberg}\ \emph {et~al.}(2013)\citenamefont
  {Rosenberg}, \citenamefont {Hultmark}, \citenamefont {Vallikivi},
  \citenamefont {Bailey},\ and\ \citenamefont {Smits}}]{RHVBS13}%
  \BibitemOpen
  \bibfield  {author} {\bibinfo {author} {\bibfnamefont {B.~J.}\ \bibnamefont
  {Rosenberg}}, \bibinfo {author} {\bibfnamefont {M.}~\bibnamefont {Hultmark}},
  \bibinfo {author} {\bibfnamefont {M.}~\bibnamefont {Vallikivi}}, \bibinfo
  {author} {\bibfnamefont {S.~C.~C.}\ \bibnamefont {Bailey}}, \ and\ \bibinfo
  {author} {\bibfnamefont {A.~J.}\ \bibnamefont {Smits}},\ }\href@noop {}
  {\bibfield  {journal} {\bibinfo  {journal} {J. Fluid Mech.}\ }\textbf
  {\bibinfo {volume} {731}},\ \bibinfo {pages} {46} (\bibinfo {year}
  {2013})}\BibitemShut {NoStop}%
\bibitem [{\citenamefont {Sano}\ \emph {et~al.}(1989)\citenamefont {Sano},
  \citenamefont {Wu},\ and\ \citenamefont {Libchaber}}]{SWL89}%
  \BibitemOpen
  \bibfield  {author} {\bibinfo {author} {\bibfnamefont {M.}~\bibnamefont
  {Sano}}, \bibinfo {author} {\bibfnamefont {X.~Z.}\ \bibnamefont {Wu}}, \ and\
  \bibinfo {author} {\bibfnamefont {A.}~\bibnamefont {Libchaber}},\ }\href@noop
  {} {\bibfield  {journal} {\bibinfo  {journal} {Phys. Rev. A}\ }\textbf
  {\bibinfo {volume} {40}},\ \bibinfo {pages} {6421} (\bibinfo {year}
  {1989})}\BibitemShut {NoStop}%
\bibitem [{\citenamefont {Procaccia}\ \emph {et~al.}(1991)\citenamefont
  {Procaccia}, \citenamefont {Ching}, \citenamefont {Constantin}, \citenamefont
  {Kadanoff}, \citenamefont {Libchaber},\ and\ \citenamefont {Wu}}]{PCCKLW91}%
  \BibitemOpen
  \bibfield  {author} {\bibinfo {author} {\bibfnamefont {I.}~\bibnamefont
  {Procaccia}}, \bibinfo {author} {\bibfnamefont {E.~S.~C.}\ \bibnamefont
  {Ching}}, \bibinfo {author} {\bibfnamefont {P.}~\bibnamefont {Constantin}},
  \bibinfo {author} {\bibfnamefont {L.~P.}\ \bibnamefont {Kadanoff}}, \bibinfo
  {author} {\bibfnamefont {A.}~\bibnamefont {Libchaber}}, \ and\ \bibinfo
  {author} {\bibfnamefont {X.~Z.}\ \bibnamefont {Wu}},\ }\href@noop {}
  {\bibfield  {journal} {\bibinfo  {journal} {Phys. Rev. A}\ }\textbf {\bibinfo
  {volume} {44}},\ \bibinfo {pages} {8091} (\bibinfo {year}
  {1991})}\BibitemShut {NoStop}%
\bibitem [{\citenamefont {Cioni}\ \emph {et~al.}(1995)\citenamefont {Cioni},
  \citenamefont {Ciliberto},\ and\ \citenamefont {Sommeria}}]{CCS95}%
  \BibitemOpen
  \bibfield  {author} {\bibinfo {author} {\bibfnamefont {S.}~\bibnamefont
  {Cioni}}, \bibinfo {author} {\bibfnamefont {S.}~\bibnamefont {Ciliberto}}, \
  and\ \bibinfo {author} {\bibfnamefont {J.}~\bibnamefont {Sommeria}},\
  }\href@noop {} {\bibfield  {journal} {\bibinfo  {journal} {Europhys. Lett.}\
  }\textbf {\bibinfo {volume} {32}},\ \bibinfo {pages} {413} (\bibinfo {year}
  {1995})}\BibitemShut {NoStop}%
\bibitem [{\citenamefont {Zhou}\ and\ \citenamefont {Xia}(2001)}]{ZX01}%
  \BibitemOpen
  \bibfield  {author} {\bibinfo {author} {\bibfnamefont {S.~Q.}\ \bibnamefont
  {Zhou}}\ and\ \bibinfo {author} {\bibfnamefont {K.-Q.}\ \bibnamefont {Xia}},\
  }\href@noop {} {\bibfield  {journal} {\bibinfo  {journal} {Phys. Rev. Lett.}\
  }\textbf {\bibinfo {volume} {87}},\ \bibinfo {pages} {064501} (\bibinfo
  {year} {2001})}\BibitemShut {NoStop}%
\bibitem [{\citenamefont {Castaing}(1990)}]{Ca90}%
  \BibitemOpen
  \bibfield  {author} {\bibinfo {author} {\bibfnamefont {B.}~\bibnamefont
  {Castaing}},\ }\href@noop {} {\bibfield  {journal} {\bibinfo  {journal}
  {Phys. Rev. Lett.}\ }\textbf {\bibinfo {volume} {65}},\ \bibinfo {pages}
  {3209} (\bibinfo {year} {1990})}\BibitemShut {NoStop}%
\bibitem [{\citenamefont {L'vov}(1991)}]{Lv91}%
  \BibitemOpen
  \bibfield  {author} {\bibinfo {author} {\bibfnamefont {V.~S.}\ \bibnamefont
  {L'vov}},\ }\href@noop {} {\bibfield  {journal} {\bibinfo  {journal} {Phys.
  Rev. Lett.}\ }\textbf {\bibinfo {volume} {67}},\ \bibinfo {pages} {687}
  (\bibinfo {year} {1991})}\BibitemShut {NoStop}%
\bibitem [{\citenamefont {Lohse}(1994)}]{Lo94}%
  \BibitemOpen
  \bibfield  {author} {\bibinfo {author} {\bibfnamefont {D.}~\bibnamefont
  {Lohse}},\ }\href@noop {} {\bibfield  {journal} {\bibinfo  {journal} {Phys.
  Rev. Lett.}\ }\textbf {\bibinfo {volume} {73}},\ \bibinfo {pages} {3223}
  (\bibinfo {year} {1994})}\BibitemShut {NoStop}%
\bibitem [{\citenamefont {Niemela}\ \emph {et~al.}(2000)\citenamefont
  {Niemela}, \citenamefont {Skrbek}, \citenamefont {Sreenivasan},\ and\
  \citenamefont {Donnelly}}]{NSSD00}%
  \BibitemOpen
  \bibfield  {author} {\bibinfo {author} {\bibfnamefont {J.~J.}\ \bibnamefont
  {Niemela}}, \bibinfo {author} {\bibfnamefont {L.}~\bibnamefont {Skrbek}},
  \bibinfo {author} {\bibfnamefont {K.~R.}\ \bibnamefont {Sreenivasan}}, \ and\
  \bibinfo {author} {\bibfnamefont {R.}~\bibnamefont {Donnelly}},\ }\href@noop
  {} {\bibfield  {journal} {\bibinfo  {journal} {Nature}\ }\textbf {\bibinfo
  {volume} {404}},\ \bibinfo {pages} {837} (\bibinfo {year}
  {2000})}\BibitemShut {NoStop}%
\end{thebibliography}

%

\end{document}